# X-ray Fractional Orbital Angular Momentum from Coherent Magnetic Scattering

P. D. Montgomery[1,*], J. S. Woods[2], M. R. McCarter[1,4], R. Divan[3], D. Czaplewski[3], W.-K. Kwok[2], U. Welp[2], R. V. Chopdekar[4], S. Roy[4], A. Barbour[6], C. Mazzoli[6], L. E. De Long[5], and J. T. Hastings[1,5]

[1]Electrical and Computer Engineering, University of Kentucky, Lexington, KY 40506, USA
[2]Materials Science Division, Argonne National Laboratory, Lemont, IL 60439, USA
[3]Center for Nanoscale Materials, Argonne National Laboratory, Lemont, IL 60439, USA
[4]Advanced Light Source, Lawrence Berkeley National Laboratory, Berkeley, CA 94720, USA
[5]Department of Physics and Astronomy, University of Kentucky, Lexington, KY 40506, USA
[6]National Synchrotron Light Source II, Brookhaven National Laboratory, Upton, NY 11973, USA
[*]patrick.montgomery@uky.edu

**Abstract.** Artificial spin ice (ASI) based on a square lattice with a topological defect are known to generate orbital angular momentum (OAM) in diffracted X-ray beams. A previous investigation of ASI with even-charge topological defects showed both charge and magnetic X-ray scattering yield photon OAM, but these were confined to integer OAM values. However, the period of the square ASI's antiferromagnetic ground state is twice the period of the structural ground state, which should lead to fractional OAM from magnetic scattering when the topological defect has odd-charge. We employed photoemission electron microscopy to confirm that these ASIs order into antiferromagnetic ground states with protected superdomain walls that provide the phase discontinuity required for fractional OAM. Resonant, coherent X-ray scattering from ASIs with topological defects of charge 1 yields integer-valued X-ray OAM at structural charge peaks and fractional X-ray OAM at magnetic peaks. For thermally active ASIs, the fractional OAM beam exhibits fluctuations in the position of the phase discontinuity and thus dynamic rotation of the beam.

## INTRODUCTION

The study of fractional photon orbital angular momentum was initiated by the experimental work of Beijersbergen et al.[1], early theoretical treatments of Franke-Arnold et al. regarding entanglement[2], and Berry's detailed study of fractional vortex beams[3]. Since then, development of fractional OAM has been driven by both fundamental considerations (e.g., high-order entanglement [4] and optical exploration of the Hilbert hotel paradox[5]), as well as applications such as optical manipulation[6,7], multiplexed communications[8], imaging[9], and sensing[10]. Recently, the study of fractional OAM was extended to the extreme ultraviolet spectral region using high harmonic generation[11]. Here, we present what we believe to be the first observation of fractional OAM in the soft X-ray region and show how it can arise from magnetic scattering from an array of nanomagnets, an artificial spin ice (ASI), as illustrated in Fig. 1.

Fractional OAM can be generated by a number of means including spiral phase-plates[1,12], fractional forked gratings[13,14], fractional spiral zone plates[15], and metasurfaces[16-18]. Phase plates are challenging to fabricate in the soft X-ray region because of the limited materials choice for phase shifters and their high absorption. Metasurfaces are perhaps even more daunting because they require patterned sub-wavelength features, which are currently impossible at X-ray wavelengths. Free standing gratings and zone plates are certainly feasible and have been successfully used for X-ray integer OAM generation[19-21]. However, here we present a novel approach in which coherent, resonant magnetic X-ray scattering produces fractional OAM based on the magnetic texture of an ASI, rather than a structural optical component. This opens the possibility of dynamic and controllable X-ray fractional OAM.

The magnetic system of interest is an array of nanomagnets, a square ASI, with a topological defect of charge 1 as shown in Fig. 1. The term ASI arises from studies of such arrays in which magnetic frustration can mimic frustration in water ice[22,23]. However, for this work, the important aspect of the square ASI is that it orders into an antiferromagnetic (AFM) ground state upon cooling below its Néel temperature, which can be near room temperature and far below the Curie temperature of the material from which it is made[24,25]. We previously showed that a ~3 nm thick $Ni_{0.8}Fe_{0.2}$ (permalloy) square ASI with a structural defect of topological charge $\mathbb{Z}_s = 2$ orders into a single-domain AFM ground state with a magnetic defect of topological charge $\mathbb{Z}_m = 1$. For an appropriate ASI geometry, this ordering occurs near room temperature and generates *odd integer-order* X-ray OAM by magnetic scattering[26]. Here, we study square ASI with a structural topological charge of $\mathbb{Z}_s = 1$ which produces magnetic defects with a fractional charge $\mathbb{Z}_m = 1/2$. An X-ray magnetic circular dichroism photoemission electron microscopy (XMCD-PEEM) image of this structure in the AFM ground state is shown in Fig. 1(c) with the magnetic texture depicted in Fig. 1(d).

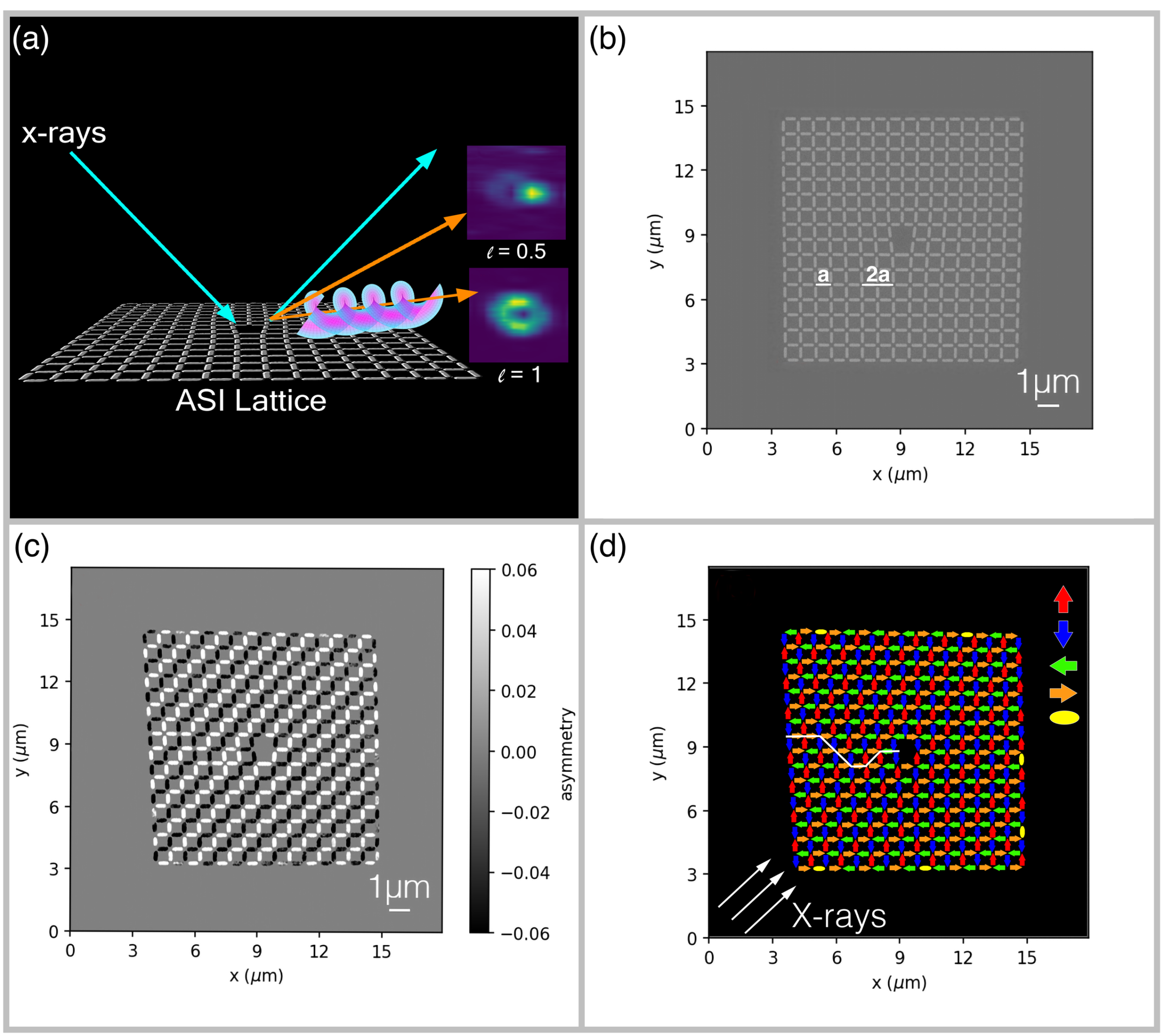

**Fig. 1: (a) Schematic of X-ray scattering from an artificial spin ice (ASI) with a topological defect of charge 1. X-rays are diffracted from the ASI to produce fractional orbital angular momentum ($\ell = 0.5$ shown) and integer-order OAM ($\ell = 1$ shown). The magnetic texture produces the fractional OAM, while the structure produces the integer-order OAM. Angles are not to scale. (b) Scanning electron micrograph (SEM) of a permalloy ($Ni_{0.8}Fe_{0.2}$) ASI with topological defect of charge 1. The defect induces OAM in the diffracted X-rays through an azimuthal phase progression of $2\pi$ radians. The structure has a lattice constant of *a* while the antiferromagnetic texture has a periodicity of 2*a*. (c) XMCD-PEEM image of a square ASI with a topological defect of charge 1 acquired at 340K. The dark and light segments show the magnetic asymmetry and reveal the AFM order. The left side of the pattern shows a protected superdomain wall. (d) Magnetic texture of the islands shown in (c). The direction of the incident X-ray beam for XMCD-PEEM is shown in the bottom left of the image. The colored arrows in the top right corner indicate the magnetic orientation of the permalloy islands. The yellow segments represent islands that have an undetermined state along the edge of the pattern due to thermal fluctuations. The indeterminate islands on the edge of the pattern have a negligible effect on the diffraction pattern from magnetic scattering. The superdomain wall is highlighted by the white line overlaying the pattern.**

Antiferromagnets with edge dislocations, defects of topological charge $\mathbb{Z}_s = 1$, were originally described theoretically by Dzyaloshkinskiĭ[27]. He recognized that an edge dislocation in an antiferromagnet requires the formation of a domain wall pinned to the topological defect. Much more recently, Drisko et al. [28]showed that a 23 nm thick $FePd_3$ square ASI with an edge dislocation does, in fact, support a protected superdomain wall terminating on topological defects and/or boundaries after annealing at $\approx 440\mathrm{K}$. Here, we find a similar structure of $Ni_{0.8}Fe_{0.2}$, when thinned to 2 to 3 nm, also yields a superdomain wall in the AFM order as shown in Fig. 1(c,d). We show that this magnetic texture produces both fractional and dynamic X-ray OAM near room temperature.

The origin of fractional OAM can be understood from the following considerations. The lattice sum for a square lattice with a topological defect can be approximated near a reciprocal lattice point, $\overrightarrow{Q_0}$, by [26,29]

$$L'(\rho', \phi') = i\,\ell\,\exp[i\ell(\phi')]\,U(\ell, \rho', \phi') \qquad (1)$$

where $\rho'$ and $\phi'$ are the local radial and azimuthal coordinates centered at $\overrightarrow{Q_0}$, $\ell$ is the OAM quantum number, and $U$ is a purely real function describing the amplitude. The OAM quantum number is given by

$$\ell = \frac{\overrightarrow{Q_0}\cdot\vec{t}}{2\pi} \qquad (2)$$

where $\vec{t}$ is the Burgers vector describing the dislocation. For a topological defect of charge 1, as shown in Fig. 1(b), the Burgers vector is simply $\vec{t} = a\hat{x}$ representing the introduction of a single structural period along the y-axis ($\mathbb{Z}_s = 1$). Thus, at the structural Bragg condition, $H = \frac{Q_x}{2\pi} = 1, 2, 3, \ldots$, we find $\ell = 1, 2, 3, \ldots$, indicating that the charge scattered beams carry integer OAM ($\hbar$, $2\hbar$, $3\hbar$, ...). This is essentially the same effect as when a conventional forked grating is used to generate OAM.

However, the magnetic case is significantly more interesting. Because the diffraction from the AFM configuration occurs at half-integer values of H and K, one immediately recognizes that $\ell$ might take on half-integer values and that the photon OAM could be fractional. This requires an abrupt $\pi$ phase discontinuity, which is present in the AFM order of the magnetic ground state because of the superdomain wall. The topological defect in the magnetic texture has charge $\mathbb{Z}_m = \frac{1}{2}$, and X-rays diffracted at the magnetic Bragg condition, $H = \frac{Q_x}{2\pi} = \frac{1}{2}, \frac{3}{2}, \frac{5}{2}, \ldots$ will yield beams with $\ell = \frac{1}{2}, \frac{3}{2}, \frac{5}{2}, \ldots$ leading to the expectation of fractional orbital angular momentum. However, the Burgers vector location and orientation is no longer determined by the structure, but rather by the superdomain wall that forms in the magnetic texture. This introduces an offset in the phase term and azimuthal dependence in the amplitude term of equation (1), which will be apparent in the experimental results.

## RESULTS

The square ASI samples used in this study utilized thin (2 to 3 nm) permalloy patterns to achieve thermal activity just above room temperature [23]. The magnetic texture was imaged using XMCD-PEEM after heating the sample in-situ to 370 K and cooling to 300 K. Measurements were performed on PEEM-3 at beamline 11.0.1.1 of the Advanced Light Source at Lawrence Berkeley National Laboratory. The resulting magnetic images confirmed that the sample equilibrated into a single AFM domain with a protected superdomain wall as shown in Fig. 1(c). The superdomain wall introduces the phase discontinuity in the magnetic scattering required for fractional OAM. The PEEM images identify the magnetic orientation of each island, as shown in Fig. 1(d), and identify the elevated energy vertices composing the superdomain wall.

Coherent, resonant magnetic X-ray scattering was conducted in a reflection geometry ($2\theta = 18°$) at the Fe $L_3$ edge (708 eV) using RCP and LCP. The polarization(s) used for fitting at each temperature can be found in Table S3 of the supplemental information. Measurements were performed on CSX (beamline 23-ID-1) at NSLS-II at Brookhaven National Laboratory. As shown in Fig. 2, magnetic scattering appears at the AFM Bragg condition (half-integer values of H and K) while charge scattering dominates at the Bragg condition for the structural lattice (integer values of H and K). Magnetic scattering gives rise to asymmetric diffraction patterns consistent with fractional OAM, while charge scattering yields nearly symmetric “donut” beams consistent with integer OAM. The asymmetric, half-integer diffraction patterns can be fit with a fractional OAM model to quantify the fractional OAM and the orientation of the phase discontinuity/superdomain wall. This model can also be used to fit the integer-order diffraction patterns, but it loses the phase information related to the superdomain wall position because of the symmetric nature of integer-order OAM.

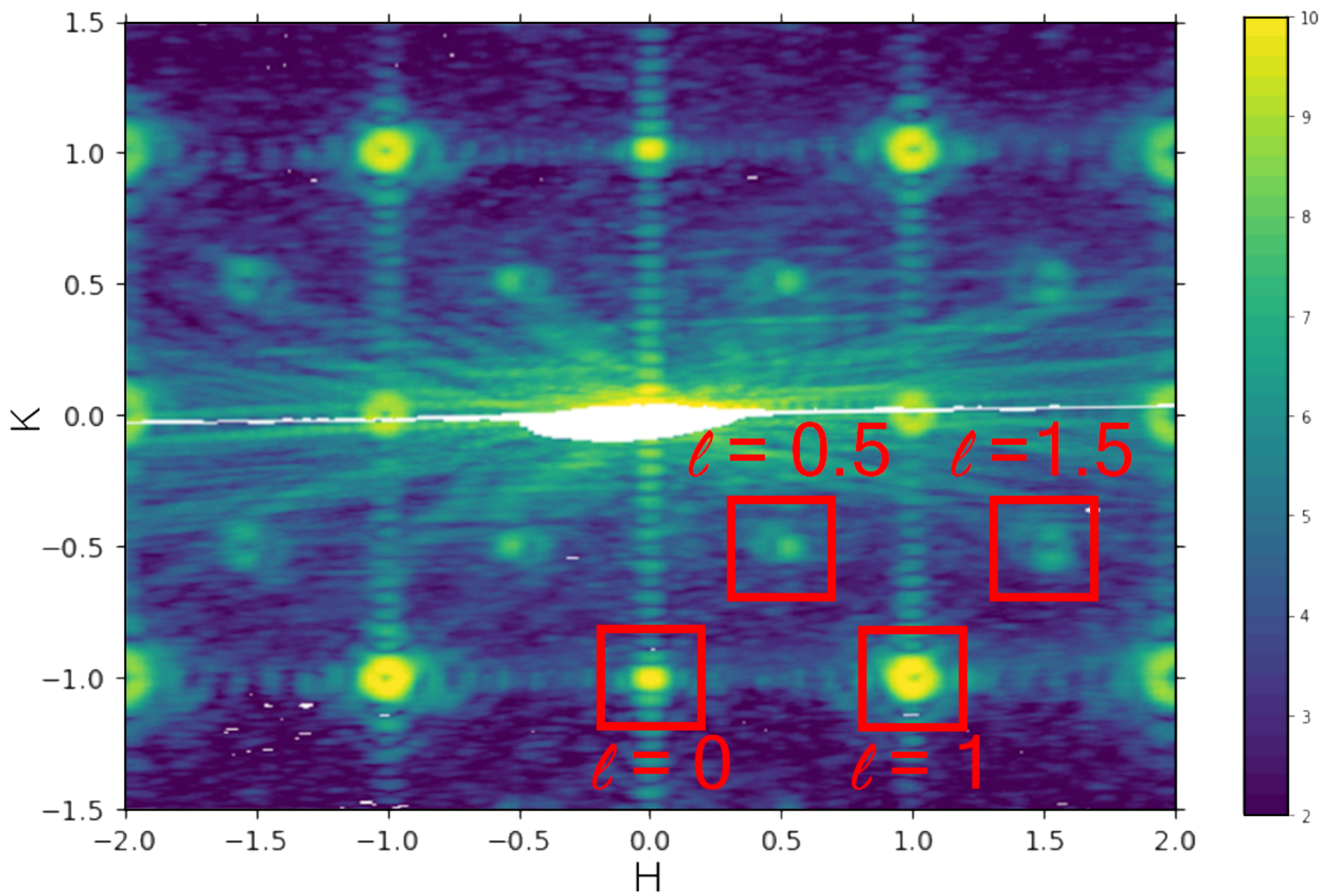

**Fig. 2: Coherent, resonant X-ray scattering from a square ASI containing a topological defect of charge 1 at a temperature of 300 K. Diffraction patterns consistent with fractional OAM can be seen at the AFM Bragg condition (half-integer H and K). Diffraction patterns consistent with integer OAM appear at the structural Bragg condition (integer H and K). Red boxes highlight the regions used to quantify the OAM. The color bar represents relative intensity on a logarithmic scale.**

## OAM of diffracted X-ray beams

Fractional OAM can be modeled as a superposition of integer Laguerre-Gauss (LG) modes given by[20,30]:

$$\psi_{LG_\ell}(\boldsymbol{r}, \varphi, z, \ell) = \sqrt{\frac{2p!}{\pi(p+|\ell|)!}}\frac{1}{w(z)}\left(\frac{r\sqrt{2}}{w(z)}\right)^{|\ell|} e^{-\frac{r^2}{w^2(z)}} L_p^{|l|}\left(\frac{2r^2}{w^2(z)}\right) e^{-ik\frac{r^2}{2R(z)}} e^{-i\ell\varphi} e^{i\psi(z)} \tag{3}$$

where $r$ is the radial coordinate, $\varphi$ is the azimuthal coordinate, $z$ is the distance from the detector to the beam waist, $\ell$ is the azimuthal index of the mode, $p$ is the radial index of the mode, $w(z)$ is the beam width at a distance z from the beam waist, $L_p^{|l|}$ is the generalized Laguerre polynomial, $k$ is the wavenumber, $R(z)$ is the radius of curvature at distance z from the beam waist, and $\psi(z)$ is the Gouy phase shift.

Integer LG modes diffracted at the structural Bragg condition can be described by equation (3) alone; however, describing fractional OAM beams at the magnetic Bragg condition requires a superposition of integer OAM modes. Berry showed that the superposition of modes required for fractional OAM is determined by the Fourier series[3]:

$$e^{i\alpha\varphi} = \frac{e^{i\pi\alpha}\sin(\pi\alpha)}{\pi}\sum_{-\infty}^{\infty}\frac{e^{i\ell\varphi}}{\alpha-\ell} \tag{4}$$

where $\alpha$ is the fractional topological charge of the OAM beam. Therefore[3],

$$\psi_{LG_\alpha}(\boldsymbol{r},\varphi,z,\alpha) = \frac{e^{i\pi\alpha}\sin(\pi\alpha)}{\pi}\sum_{\ell=-\infty}^{\ell=\infty}\frac{\psi_{LG_\ell}(\boldsymbol{r},\varphi,z,\ell)}{\alpha-\ell} \tag{5}$$

We fit absolute square of equation (5), $\left|\psi_{LG_\alpha}\right|^2$ , to the experimentally measured intensity of the scattered X-ray beams. The asymmetry of fractional OAM beams allows us to reliably fit $\alpha$ using only the intensity from magnetic scattering. When fitting fractional LG beams to the experimental intensity the first term of equation (5) becomes $\frac{\sin^2(\pi\alpha)}{\pi^2}$. To simplify the fit, we can combine this first term with a relative amplitude, $A$, as shown in equation (6), which is used to fit the diffraction pattern of the scattered X-rays. We constrained the superposition to $-20 < \ell < 20$ after verifying that $\alpha$ values converged within this range and that modal contributions of each $|\ell| > 20$ were < 0.5%. With these changes, we can simplify equation (5) as follows:

$$\psi_{LG_\alpha}(\boldsymbol{r},\varphi,z,\alpha) = A\sum_{\ell=-20}^{\ell=20}\frac{\psi_{LG_\ell}(\boldsymbol{r},\varphi,z,\ell)}{\alpha-\ell} \tag{6}$$

This equation can be used to approximate an LG beam for fractional modes when $\alpha$ is not an integer.

The denominator of the term inside of the summation of equation (6), $\frac{\psi_{LG_\ell}(\boldsymbol{r},\varphi,z,\ell)}{\alpha-\ell}$, results in a division by zero when $\alpha$ is an integer. To avoid numerical errors in equations (5, 6), we set $\alpha = \alpha + 10^{-12}$ when fitting integer peaks. This allows the same equation to be used for fractional and integer values of $\alpha$. Additional details regarding the fitting algorithm and parameters fit can be found in the supplemental information.

The variable parameters used in our fitting algorithm included the beam center on the detector ($x_0$ and $y_0$), beam waist ($w_0$), phase rotation offset ($\varphi_0$), intensity ($A_0$), and the fractional OAM ($\alpha_0$). The experimental diffraction pattern of each fractional and integer mode was extracted from the scattering intensity as shown in Fig. 2. Both magnetic peaks with $\ell \approx \frac{1}{2}, \frac{3}{2}$ and the charge peaks with $\ell \approx 0, 1$ were fit using equation (6) and 600 sequential frames per $\ell$ mode. The set of 600 frames were divided into 20 subsets, each containing 30 sequential frames. Each of these subsets were then averaged to create 20 frames, which were then fit individually. The fitted parameters for each of the 20 frames were then averaged, and a simulated diffraction pattern was calculated using $r = \sqrt{(x-x_0)^2 + (y-y_0)^2}$, $\varphi = \arctan\left(\frac{y-y_0}{x-x_0}\right) - \varphi_0$, $w_{fit} = w_{expected} - w_0$, $\ell = \alpha_{expected} - \alpha_0$, and $A = A_{expected} - A_0$ in equation (3), where:

$$w(z) = w_{fit}\sqrt{1 + \left(\frac{z\lambda}{\pi w_{fit}^2}\right)^2} \tag{7}$$

The experimental diffraction intensity patterns are compared with the calculated diffraction intensity patterns for $\ell = 0, \frac{1}{2}, 1,$ and $\frac{3}{2}$ in Fig. 3. The calculated diffraction patterns closely match the experimental diffraction patterns. In addition, the peak and relative intensities are well-modeled in all cases except $\ell = 0$, where the average $\alpha$ closely approaches zero leading to numerical divergence that is not present in any of the individual fitted $\alpha$ values. This is discussed in detail in Section 3 of the supplementary information and does not affect the conclusion that the $\ell = 0$ peak is consistent with absence of OAM.

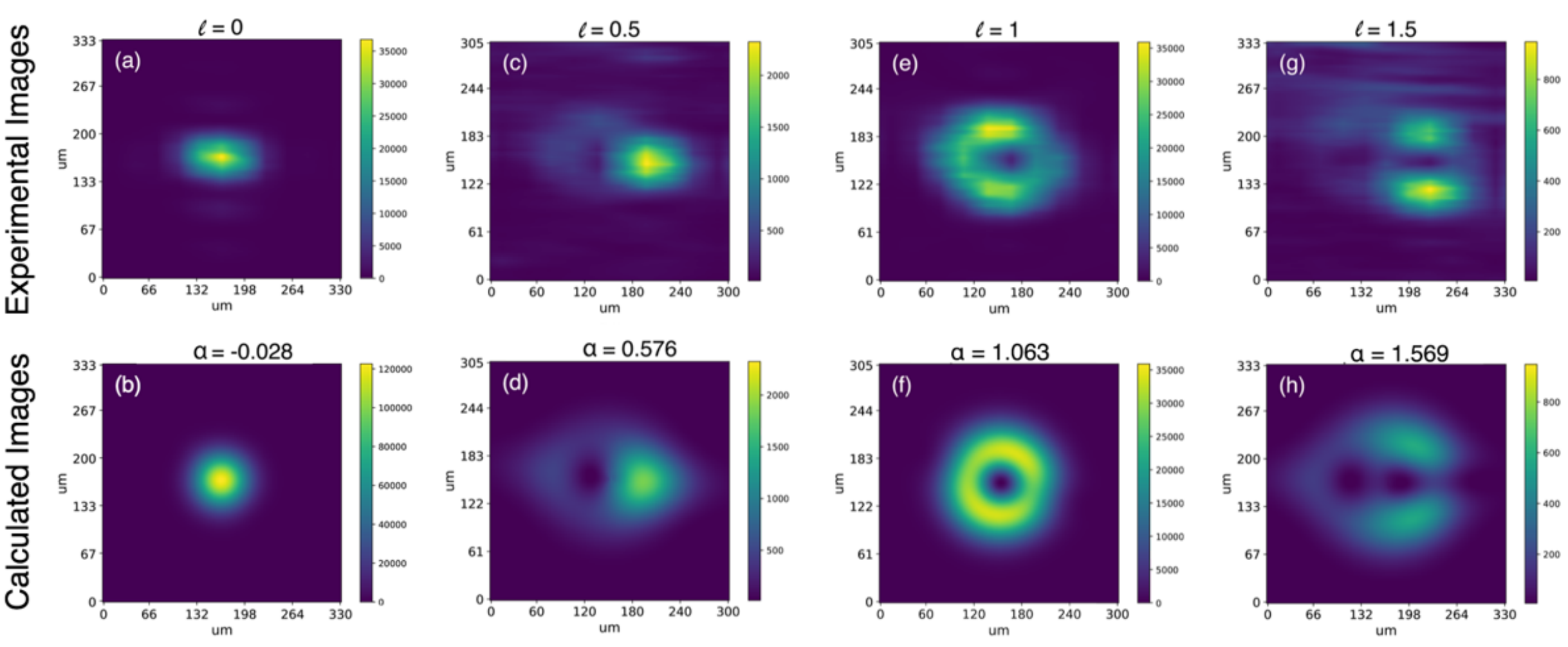

**Fig. 3: Experimental and calculated magnetic scattering for OAM beams diffracted from a $\mathbb{Z}_s = 1$ ASI at 300 K. Experimental images are labeled with the expected $\ell$ values. Images calculated from the fitted parameters are labeled with average $\alpha$ values. Good agreement between expected and measured OAM values was found for all four patterns. The color scale represents relative intensity.**

The fit of the fractional OAM, $\alpha$, was compared to the expected OAM mode, $\ell$, for $\ell = 0, \frac{1}{2}, 1, \frac{3}{2}$ at temperatures of 300 - 330 K as shown in Fig. 4. Our results show that the diffraction pattern obtained from magnetic scattering at the half-order Bragg condition is consistent with fractional OAM at temperatures of 300 – 330 K. The agreement between measured and expected values for the topological charge, $\alpha$, are very good. The expected integer charges fall within the error bounds in all cases except for $\ell = 1$ at 300 K, where the lower bound is within 2.3% of the expected value. The expected fractional charges tend to fall slightly outside the error bounds of the calculated values. This small discrepancy could result from the formation of a more complex superdomain wall that does not extend in a straight line from the patterned defect to the edge of the ASI. Regardless, fractional OAM is clearly generated as further evidenced by the modal decomposition discussed in the following section and shown in Fig. 5.

## Modal decomposition of LG beams with fractional OAM

Fractional OAM can be modeled by a superposition of multiple $\ell$ modes as described in equation (6). The difference in integer and fractional OAM beam modal composition can be seen in Fig. 5, which shows the superposition decomposition of $\ell$ modes fit at 300 K. The magnetic peak at $H = 1/2, K = -1/2$, with expected $\ell = 1/2$ and fitted value of $\alpha = 0.576 \pm 0.002$, shows the largest contributions from the nearest integer modes ($\ell = 0 \text{ and } 1$) with declining contributions for other $\ell$ modes. The magnetic peak at $H = 3/2, K = -1/2$, with expected $\ell = 3/2$ and fitted value of $\alpha = 1.569 \pm 0.013$, shows the largest contribution from the $\ell = 1 \text{ and } 2$ modes with declining contributions from other $\ell$ modes. As expected, beams with fractional OAM are shown to be described by the superposition of integer OAM modes.

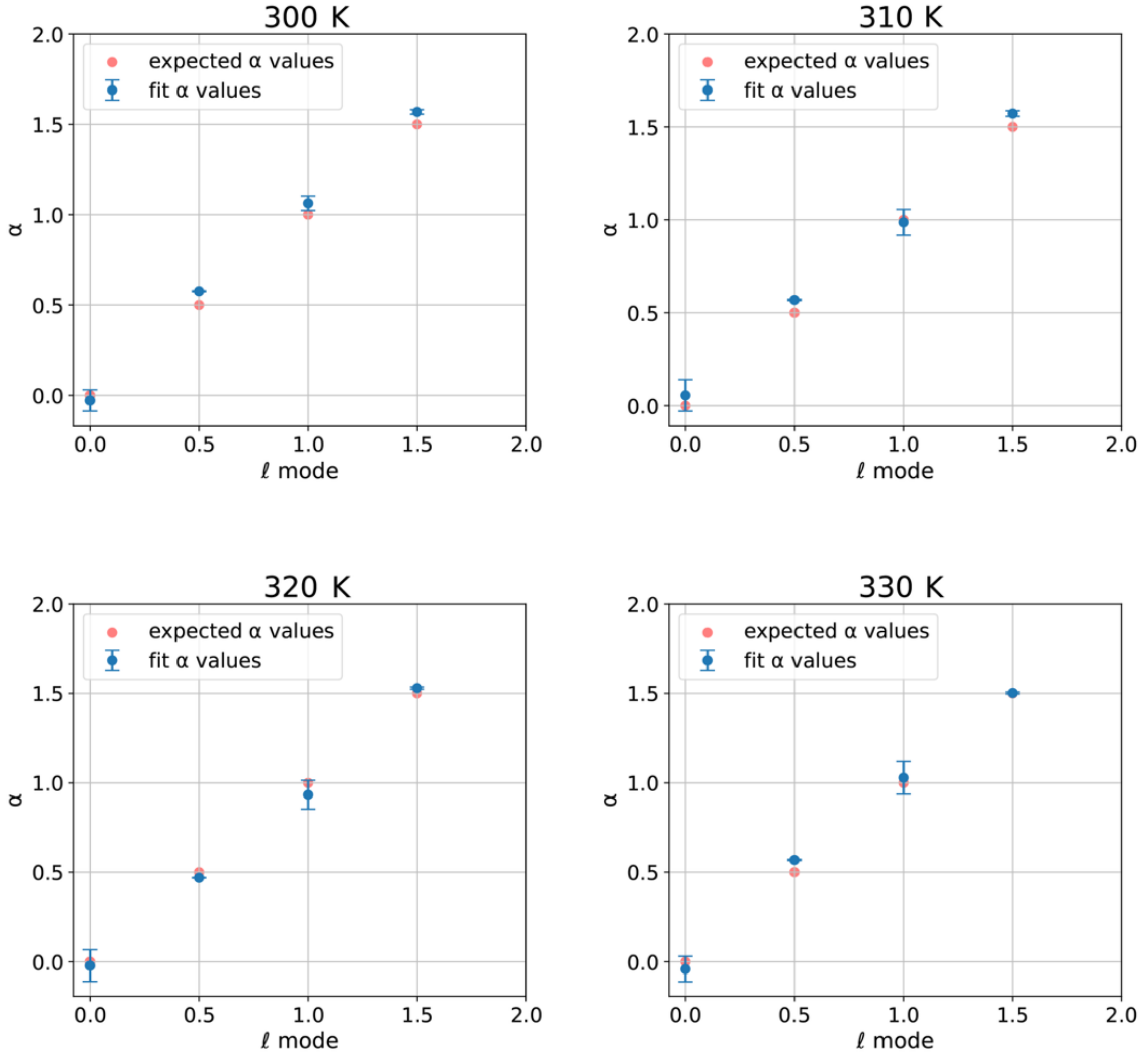

**Fig. 4. Comparison of expected $\ell$ values to the average fitted $\alpha$ values for temperatures from 300 K – 330 K. The fitted values of $\alpha$ for magnetic scattering ($\ell = 1/2\,,3/2$) have a low degree of error, and the results are consistent with fractional OAM. The $\alpha$ values remain consistent across this temperature range where the ASI magnetic configuration is largely static. Error bars represent the standard deviation of the fitted value of $\alpha$ for each $\alpha_{expected} = \ell$.**

In contrast, the charge peak at $H = 0, K = -1$, with expected $\ell = 0$ and fitted value of $\alpha = -0.028 \pm 0.059$ is primarily composed of the $\ell = 0$ mode, with small contributions from additional modes. Likewise, the charge peak at $H = 1, K = -1$, with expected $\ell = 1$ and fitted value of $\alpha = 1.063 \pm 0.040$ is primarily composed of the $\ell$ =1 mode, with small contributions from additional modes. Thus, the beams expected to carry integer OAM are found to be well modeled by a single LG mode.

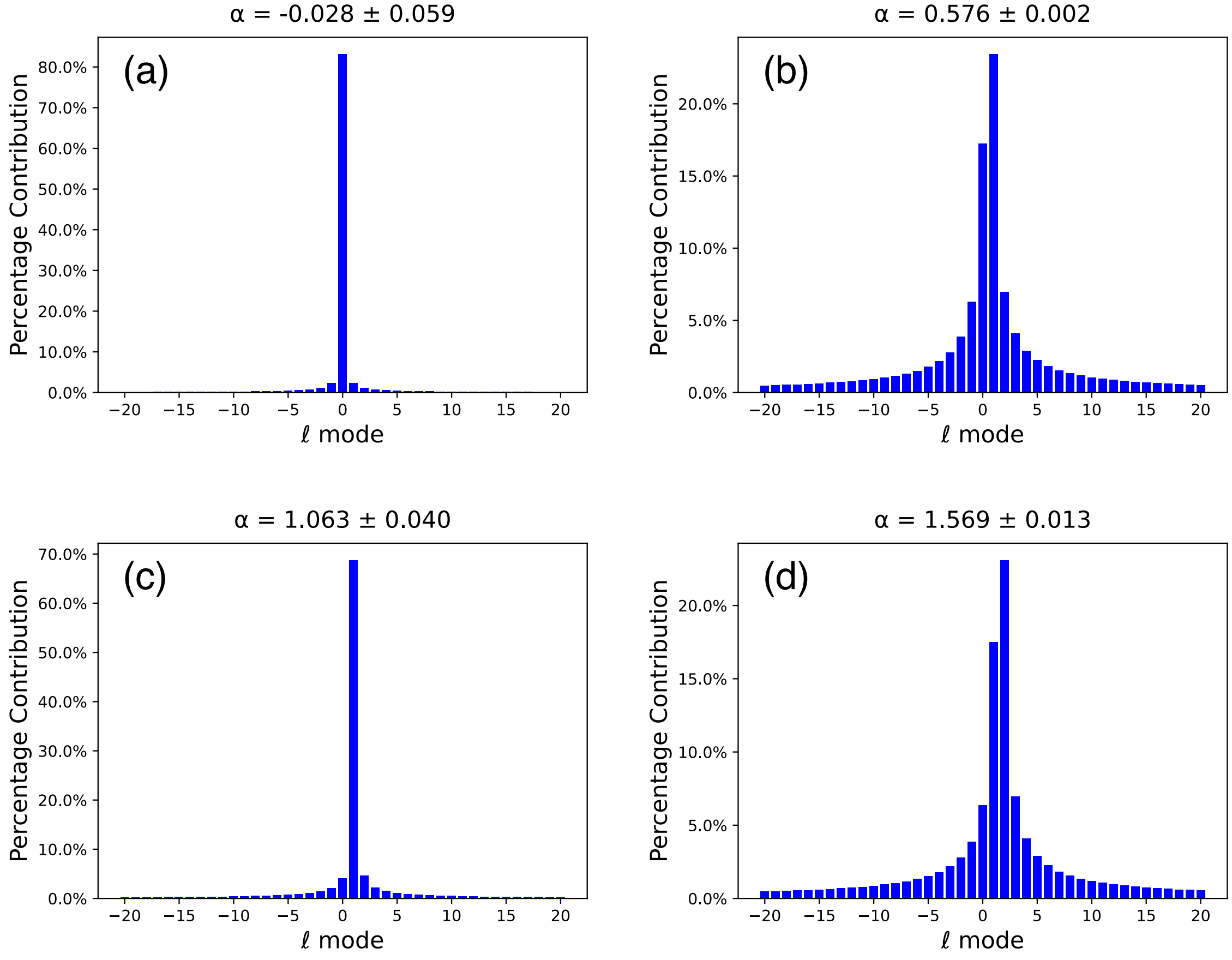

**Fig. 5. Modal decomposition of magnetic and charge peaks at 300 K. (a) The modal decomposition of $\psi_{LG_\alpha}(\alpha_{fit})$ for the $\ell = 0$ charge peak. (b) The modal decomposition of $\psi_{LG_\alpha}(\alpha_{fit})$ for the $\ell = 1/2$ magnetic peak. (c) The modal decomposition of $\psi_{LG_\alpha}(\alpha_{fit})$ for the $\ell$ =1 charge peak (d) The modal decomposition of $\psi_{LG_\alpha}(\alpha_{fit})$ for the $\ell = 3/2$ magnetic peak. The charge peaks ($\ell = 0, 1$), are primarily composed of the expected $\ell$ modes with only small contributions from higher order modes. The magnetic peak with expected $\ell = 1/2$ is primarily composed of a superposition of $\ell = 0, 1$ modes while the magnetic peak with expected $\ell = 3/2$ is primarily composed of a superposition of $\ell = 1, 2$ modes. However, both fractional OAM beams require the superposition of a broad range of higher $\ell$ modes evidenced by decaying contributions of additional $\ell$ modes.**

## Dynamic rotation of fractional OAM beams

The phase rotation, $\varphi$, is related to the azimuthal position of the superdomain wall and can be used to track fluctuations in the magnetic texture at higher temperatures. At temperatures of 300 K, 310 K, 320 K, and 330 K, the domain wall was found to be stable throughout the 180 s duration of data acquisition for right and left circular polarizations (RCP and LCP respectively). This resulted in the expected $\ell = 1/2$ mode showing negligible phase rotation for both RCP and LCP at 300 K, 310 K, and 320 K. At 330 K, the phase rotation was negligible during the RCP data acquisition, shifted between RCP and LCP measurements, and was then stable with negligible rotation during the LCP data acquisition. The polarizations used for fitting at each temperature can be found in Section 2, Table S3, of the supplemental information. Additional details regarding the fitting of $\varphi$ can be found in Section 4 of the supplemental information.

Significant fluctuations of the phase rotation were observed at 340 K in both RCP and LCP data sets. The total period of 180 s for each polarization was divided into 600 individual frames averaged over a period of 0.3 s each. The rotation of the fractional OAM beam with RCP illumination at 340 K is shown in Fig. 6. The figure shows fluctuations for the first 108 s (360 frames) of the experiment, where the angle of $\varphi$ was found to rotate up to 1.34 radians within 0.3 s, but showed a smaller average fluctuation of 0.18 radians per 0.3 s frame.

In the frames beyond 108 s, the fluctuations in $\varphi$ were larger and more frequent. Inspection of the magnetic scattering beyond 108 s revealed multiple speckles indicative of the presence of additional domain walls [25]. Further study with shorter acquisition times would be needed to determine the behavior of the domain wall fluctuations seen in frames beyond 108 s at 340 K. The fitting algorithm used in this study is unable to account for multiple speckles. Therefore, the data and fit after 108 s were excluded from Fig 6.

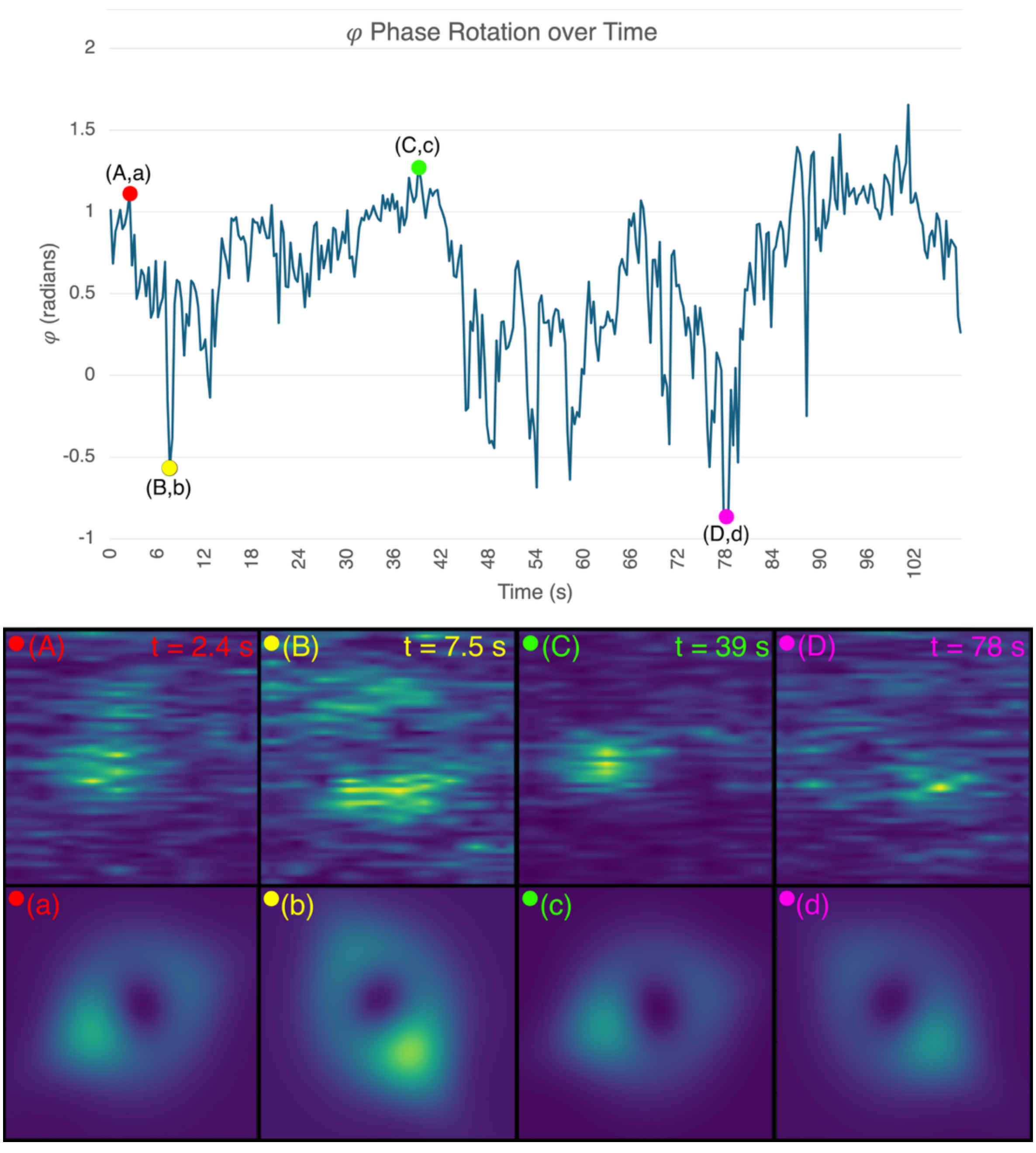

**Fig. 6: Phase rotation of phi ($\varphi$) over time at 340 K. The graph in the top half of the image shows the fitted angle of $\varphi$ over a 108 s period. The blue line shows the fitted phase rotation of phi ($\varphi$) for the $\ell$ = 0.5 mode. The colored circles on the graph correspond to experimental and calculated images at the bottom half of the figure. (A) The experimental diffraction intensity pattern at t = 2.4 s and $\varphi$ = 1.1 radians. (a) The calculated diffraction intensity pattern at t = 2.4 s and $\varphi$ = 1.1 radians generated from fit values. (B) The experimental diffraction intensity pattern at t = 7.5 s and $\varphi$ = -0.57 radians. (b) The calculated diffraction intensity pattern at t = 7.5 s and $\varphi$ = -0.57 radians generated from fit values. (C) The experimental diffraction intensity pattern at t = 39 s and $\varphi$ = 1.3 radians. (c) The calculated diffraction intensity pattern at t = 39 s and $\varphi$ = 1.3 radians generated from fit values. (D) The experimental diffraction intensity pattern at t = 78 s and $\varphi$ = -0.86 radians. (d) The calculated diffraction intensity pattern at t = 78 s and $\varphi$ = -0.86 radians generated from fit values.**

## DISCUSSION AND CONCLUSIONS

Our results show that scattering at magnetic Bragg peaks from square ASI with a topological defect of charge 1 yields fractional X-ray OAM. As a result of the asymmetric shape of fractional OAM beams, detection of the half-order diffraction intensity patterns from magnetic scattering provides strong evidence for X-ray fractional OAM. The significance of this is multifaceted. We have shown that: 1) Fractional OAM can be observed in the soft X-ray region; 2) ASIs with odd-charge topological defects can produce fractional OAM from X-ray magnetic scattering; 3) Fractional OAM from ASIs is well modeled by a superposition of multiple integer OAM modes; 4) It is possible to determine the fractional mode of X-ray magnetic scattering; and 5) The AFM superdomain wall in thin permalloy ASIs with single edge dislocations state is stable at lower temperature ranges (300 K - 320 K). At higher temperature ranges (340 K), the domain wall fluctuations are significant in the AFM state leading to dynamic, fractional OAM.

Fractional OAM produced from magnetic scattering is significant because it produces a vortex beam that can be switched on and off through applied magnetic field and temperature[26]. Magnetic scattering present while the ASI is in the AFM ground state can be removed by magnetizing the ASI in-plane or heating above the system's Néel temperature[26]. This results in fractional OAM beams that can be manipulated by a magnetic field while the integer OAM from charge scattering remains unaffected. Fractional OAM-specific applications include particle guiding and sorting[31], high-dimensional photon entanglement[2,31,32], and quantum communications[33], all of which could be considered in the X-ray region in the future.

Our results show that ASIs with odd-charge topological defects produce fractional OAM from X-ray magnetic scattering in the form of a superposition of modes. These modes can be described by Berry's model from equation (5)[3]. This leads to the possibility of fractional OAM that can be "tuned" by creating structures that produce a specific superposition of modes. Using more sophisticated structures, these superpositions could be further tuned by external magnetic fields.

We have created a algorithm that accurately fits and simulates experimental data for fractional OAM from magnetic scattering. This can be used to determine fractional OAM modes in the future and as a starting point to fine-tune ASI structures to produce specific fractional OAM modes. The approach to determining the superposition of modes using equation (6) is superior to fitting a superposition of LG beams using a summation of equation (3). The latter approach requires fitting up to 119 additional parameters for the range of modes considered in this study ($-20 \leq \ell \leq 20$) and is much more computationally intensive.

In conclusion, by connecting Berry's and Dzyaloshkinskiĭ's works[3,27], we have shown that X-ray fractional OAM is generated by magnetic scattering from odd-charge topological defects in square ASIs. The presence of a protected, but mobile, superdomain wall provides the required phase step for fractional OAM and leads to dynamic phase shifts in the OAM beams during thermal fluctuations. Finally, fractional OAM is well modeled by a superposition of integer modes with at least 41 modes contributing to the X-ray OAM beams generated here.

## METHODS

### Sample fabrication

ASI samples were created using standard e-beam lithography techniques and a liftoff procedure to fabricate the patterned magnetic thin films for XMCD-PEEM and coherent, soft X-ray scattering experiments. Samples were patterned on Si (100) substrates that are first sonicated in acetone for 120 s, then sonicated in IPA for 120 s and then rinsed in DI water to remove any organic material that may be on the substrate surface. The substrates are then spin coated with a positive resist, PMMA 495 A2 and PMMA 950 A2, creating a bilayer. The first layer is PMMA 495 A2, and is spin coated at 1,500 rpm for 50 s and then baked on a hotplate at 180 °C for 180 s. After the first layer is cured, a second layer of PMMA 950 A2 resist is spin coated at 2,500 rpm for 50 s and then baked at 180 °C for 180 s. The bilayer of PMMA resist is then exposed by e-beam lithography using a Raith e-Line or Raith 150 system. The e-beam exposes the resist in the desired pattern created using a Python algorithm and exported as a .gds file.

The exposed substrate is then developed in a mixture of ethanol and DI water at a ratio of 4:1 for 45 s, then submerged in pure ethanol for 15 s and finally rinsed in DI water for 15 s. The resist is positive, meaning that the exposed areas of the resist are dissolved and removed during development. Thus, the resist is removed in the shape of the exposed pattern leaving only the bare Si substrate. Permalloy ($Ni_{0.8}Fe_{0.2}$) and Al are then deposited on the patterned substrate using an AJA magnetron sputtering system. Samples are loaded and reach a base pressure of $1.0 \times 10^{-8}$ torr. Argon is introduced to the chamber with a flow rate of 30 sccm and the deposition pressure is set to ~3 mTorr. Permalloy is sputtered at 20 W to establish a rate of 0.9 Å/s. Al is sputtered at 15 W to establish a rate of 0.6 Å/s.

Liftoff is performed using Microposit 1165 solution. The samples are submerged in 1165 and kept at a temperature of 75 °C for ~18 hours. After ~18 hours, the sample is sonicated in 1165 solution to remove any unwanted resist and deposited metal. The sample is then sonicated in acetone for 60 s and rinsed in IPA and DI Water to clean the sample and pattern. The 1165 dissolves the remaining resist leaving behind only the deposited permalloy/Al that was in direct contact with the Si substrate. The resulting metal left behind is in the desired pattern adhered to the Si substrate. The samples are then imaged using a scanning electron microscope to confirm the viability of the pattern and that liftoff was complete.

Separate samples were prepared for PEEM and coherent X-ray scattering because the synchrotron beam times were scheduled three months apart. The use of freshly prepared samples ensured that possible oxidation occurring through or around the alumina passivation layer did not influence the experiments. In both cases, the magnetic islands were 490 nm long and 160 nm wide arranged on a square lattice with a lattice constant of 640 nm. The target permalloy thickness was 2.4 nm for the XMCD-PEEM samples and 2.0 nm for the coherent X-ray scattering samples. In both cases, the aluminum capping layer, which subsequently oxidizes in air, was deposited with a target thickness of 1.5 nm. The thinner permalloy for the X-ray scattering samples was chosen to reduce the temperature at which fluctuations began. This was observed to be the case as the fluctuations in the XMCD-PEEM sample were observed at $T \geq$ 350K, while fluctuations in the X-ray scattering sample were observed at $T \geq$ 330 K.

## X-ray photoemission electron microscopy

XMCD-PEEM experiments were conducted using the PEEM-3 end station at beamline 11.0.1.1 of the Advanced Light Source at Lawrence Berkeley National Laboratory. The X-ray energy was tuned to the Fe L3 absorption edge near 708 eV. Images were acquired with right and left circularly polarized (RCP and LCP, respectively) illumination, and the difference was taken to isolate the magnetic contribution from the elemental contribution. Samples were heated in-situ to 370 K then cooled to 300 K. XMCD-PEEM images were then acquired during heating from 340 to 380 K in 10 K steps. Images were acquired at a rate of 3 s per image with RCP and LCP X-ray excitation. The difference between images acquired with opposite polarizations was taken to emphasize the magnetic contribution.

## Coherent, resonant X-ray scattering

Coherent, resonant, soft X-ray scattering experiments were conducted at the Coherent Soft X-ray (CSX, 23-ID-1) beamline at the National Synchrotron Light Source II at Brookhaven National Laboratory. A reflection geometry was employed with a scattering angle of $2\theta = 18^{\circ}$. The energy of the incident X-ray beam was tuned to the Fe L3 edge (708 eV) to enhance magnetic sensitivity. The incident beam was passed through a $10\ \mu m$ diameter pinhole to control transverse coherence. Scattered X-ray photons were collected using a fast CCD detector positioned 340 mm away from the sample. The acquisition time was 200 ms per image and the acquisition rate was 300 ms per image. A total of 600 images were acquired at each temperature for each circular polarization. A 200 s delay was included after switching polarizations to allow additional time for thermal stabilization. A flat field image was acquired at the beginning of the experiment and dark images were acquired before each set of 600 images (frames) were taken.

## Data availability

The data that support the findings of this study are available from the corresponding author upon reasonable request.

## Acknowledgements

This work was supported by the U.S. Department of Energy, Office of Science, Office of Basic Energy Sciences under Award Number DE-SC-0024346. This research used the 23-ID-1 (CSX) beamline of the National Synchrotron Light Source II, a U.S. Department of Energy (DOE) Office of Science User Facility operated for the DOE Office of Science by Brookhaven National Laboratory under Contract No. DE-SC0012704 and resources made available through BNL/LDRD#19-013. This research used resources (beamline 11.0.1.1, PEEM-3) of the Advanced Light Source, a U.S. DOE Office of Science User Facility under contract no. DE-AC02-05CH11231. The research was supported by the U.S. Department of Energy, Office of Science, Basic Energy Sciences, Materials Sciences and Engineering Division. The use of facilities at the Center for Nanoscale Materials, an Office of Science user facilities, was supported by the U.S. Department of Energy, Basic Energy Sciences, under Contract No. DE-AC02-06CH11357. This work was performed in part at the University of Kentucky Center for Nanoscale Science and Engineering and Center for Advanced Materials, members of the National Nanotechnology Coordinated Infrastructure (NNCI), which is supported by the National Science Foundation (ECCS-2025075).

We would like to thank Dr. Roland Koch for his contributions to this work at the Advanced Light Source at Lawrence Berkeley National Laboratory.

## Author Contributions

P.D.M., J.S.W, S.R., A.B., C.M., L.E.D., and J.T.H. conceived and designed the research. J.S.W, W.K.K., U.W., R.D., and D.C. conducted sample fabrication and characterization. J.S.W and R.V.C. conducted XMCD-PEEM experiments. J.S.W., A.B., C.M., and J.T.H. conducted coherent X-ray scattering experiments. P.D.M. and M.R.M. conducted data analysis, simulation, and figure preparation. All authors contributed to the manuscript.

## Competing Interests

The authors declare no competing interests.

## Additional Information

A document containing supplemental information accompanies this manuscript.

# Supplemental Information for X-ray Fractional Orbital Angular Momentum from Coherent Magnetic Scattering

## GENERAL FITS

Sets of X-ray scattering data (600 frames each for right and left circularly polarized illumination) were loaded into a python script, which was used as the basis for all data analysis. The individual frames for $\ell = 0, \frac{1}{2}, 1, \frac{3}{2}$ were cropped (see Fig. 2, main text) to a size of 300 - 330 $\mu$m per side to maintain consistency while allowing each frame to be centered with room around the edges. Each frame from each set was resampled to 100 x 100 pixels. The cropped frames for each $\ell$ value was used in a comparison function, where the x and y coordinate offsets ($x_0$ and $y_0$), beam waist ($w_0$), phase rotation offset ($\varphi_0$), amplitude offset ($A_0$), and the fractional phase step ($\alpha$) were manually fit using equation (6). The parameter values from the manual fit were used as starting values in the fitting algorithm.

For each $\ell$ mode, the cropped frame sets were divided into 20 subsets, with each subset containing 30 consecutive frames which were then averaged to create 20 averaged frames. The corresponding averaged frames of RCP and LCP polarization were then summed at temperatures where $\varphi$ was stable between scans. For higher temperatures where $\varphi$ was unstable between scans, only one polarization was used (Table S3 describes the polarizations used at each temperature). The resulting sets of 20 frames for each $\ell$ mode were used in the fitting algorithm. Frames at 330 K exhibited a phase rotation in $\varphi$ that occurred in the time between RCP and LCP data collection, which would have led to summed frames with mismatched phase rotations, so only LCP was used at this temperature. Frames at 340 K exhibited dynamic phase rotation, so only RCP was used at this temperature.

The fitting algorithm fit the parameters $x_0$, $y_0$, $w_0$, $\varphi_0$, $A_0$, and $\alpha_0$, using equation (6) for each averaged frame. This provided 20 separate fits for each frame set. The mean and standard deviation for each parameter were calculated for each $\ell$ frame set. These values were used to simulate the fit results in a comparison function utilizing equation (3).

## PARAMETER FITTING

The result of the parameter fits for data collected at 300K, 310K, 320K, and 330K are shown in Tables S1 & S2. Parameter fits included the center offsets ($x_0$ and $y_0$), phi offset ($\varphi_0$), beam waist offset ($w_0$), amplitude offset ($A_0$), and the fractional phase step offset ($\alpha_0$). The center offset of $(x - x_0, y - y_0)$ for all frames was minimal, with a maximum of 1.11% offset in any direction in relation to frame size, indicating that our frames were well-centered as expected. The amplitude fits ($A = A_{expected} - A_0$) were as expected based on manual fits of equations (3) and (6), and the calculated range of $A$.

The beam waist fit ($w_{fit} = w_{expected} - w_0$) varied slightly based on the $\ell$ mode but was consistent within a 1.7 $\mu$m range across all $\ell$ modes. The average beam waist fit across all temperatures and modes was 3.0 $\mu$m, slightly lower than the expected value of 5 $\mu$m. The difference is likely due to the different sizes of each OAM diffraction pattern at the different $\ell$ modes. The scale of the diffraction intensity patterns of each $\ell$ mode at 300 K can be seen in Fig. 3, main text. The scale of the diffraction intensity patterns at all temperatures remained consistent for each $\ell$ mode at all temperatures which were fit. The size of the diffraction intensity pattern has a direct correlation with the beam waist fit.

The phase rotation ($\varphi = \arctan\left(\frac{y-y_0}{x-x_0}\right) - \varphi_0$) only provides meaningful results for fractional OAM. Because integer order OAM is symmetrical, the diffraction pattern will appear the same regardless of phase rotation. The fit for the phase rotation of fractional OAM was as expected. This can be seen in the alignment of the patterns of $\ell = 1/2\,, 3/2$ in Fig. 3, main text. The phase rotation of the integer $\ell$ modes were fit, but the data is not shown in Table S2 because it does not carry meaningful information for integer-order modes.

Table S1: Fit parameter values for fractional $\boldsymbol{\ell}$

| | | Fit parameter values | | | | Standard deviation of fit parameters | | | |
|---|---|---|---|---|---|---|---|---|---|
| $\ell$ | T (K) | $\alpha$ | $w$ ($\mu$m) | $\varphi$ (rad) | $A$ | $\sigma_\alpha$ | $\sigma_w$ ($\mu$m) | $\sigma_\varphi$ (rad) | $\sigma_A$ |
| 0.5 | 300 | 0.5761 | 2.6651 | 1.4765 | 0.0013 | 0.0024 | 0.0025 | 0.0060 | 0.0000 |
| 1.5 | 300 | 1.569 | 2.6086 | 1.562 | 0.0010 | 0.013 | 0.0095 | 0.031 | 0.0000 |
| 0.5 | 310 | 0.5686 | 2.3628 | 1.5282 | 0.0011 | 0.0033 | 0.0043 | 0.0074 | 0.0000 |
| 1.5 | 310 | 1.572 | 2.5376 | 1.7706 | 0.0008 | 0.015 | 0.0086 | 0.0068 | 0.0000 |
| 0.5 | 320 | 0.4692 | 2.9120 | 3.136 | 0.0006 | 0.0012 | 0.0045 | 0.029 | 0.0000 |
| 1.5 | 320 | 1.5294 | 2.3613 | 3.009 | 0.0006 | 0.0061 | 0.0046 | 0.067 | 0.0000 |
| 0.5 | 330 | 0.5682 | 2.491 | 0.688 | 0.0006 | 0.0029 | 0.028 | 0.026 | 0.0000 |
| 1.5 | 330 | 1.5021 | 2.1843 | 3.102 | 0.0005 | 0.0056 | 0.0042 | 0.024 | 0.0000 |

Table S2: Fit parameter values for integer $\boldsymbol{\ell}$

| | | Fit parameter values | | | Standard deviation of fit parameters | | |
|---|---|---|---|---|---|---|---|
| $\ell$ | T (K) | $\alpha$ | $w$ ($\mu$m) | $A$ | $\sigma_\alpha$ | $\sigma_w$ ($\mu$m) | $\sigma_A$ |
| 0 | 300 | -0.028 | 3.6922 | 0.0006 | 0.059 | 0.0030 | 0.0004 |
| 1 | 300 | 1.063 | 3.2148 | 0.0013 | 0.040 | 0.0097 | 0.0005 |
| 0 | 310 | 0.056 | 3.738 | 0.0009 | 0.085 | 0.018 | 0.0003 |
| 1 | 310 | 0.986 | 3.163 | 0.0012 | 0.069 | 0.013 | 0.0003 |
| 0 | 320 | 0.022 | 3.7715 | 0.0004 | 0.089 | 0.0071 | 0.0003 |
| 1 | 320 | 0.934 | 3.2276 | 0.0009 | 0.081 | 0.0047 | 0.0004 |
| 0 | 330 | -0.040 | 3.8981 | 0.0005 | 0.072 | 0.0040 | 0.0001 |
| 1 | 330 | 1.029 | 3.235 | 0.0011 | 0.091 | 0.022 | 0.0004 |

Table S3: Polarizations used for fitting

| Temperature (K) | Polarization Used | $\ell$ |
|---|---|---|
| 300 | RCP + LCP | 0, 0.5, 1, 1.5 |
| 310 | RCP + LCP | 0, 0.5, 1, 1.5 |
| 320 | RCP + LCP | 0, 0.5, 1, 1.5 |
| 330 | LCP | 0, 0.5, 1, 1.5 |
| 340 | RCP | 0.5 |

## $\boldsymbol{\ell = 0}$ FIT

It should be noted that the difference in the calculated and experimental intensity for $\ell = 0$ in Fig. 3 (a, b) in the main text is due to the averaging of the 20 $A_{fit}$ and $\alpha_{fit}$ values. As $\alpha \to 0$, $A$ decreases, and this is seen across all the individual fits for $\ell = 0$. The parameter of $\alpha_{fit}$ was bound between
-0.4:0.4. Because $\alpha_{fit}$ can result in a positive or negative value, the average can result in a value close to zero, as expected for the $\ell = 0$ mode. The value of $A_{fit}$ was bound to positive values to remain consistent with the $\frac{\sin^2(\pi\alpha)}{\pi^2}$ term it replaces. This resulted in a larger average $A_{fit}$ value than is reasonable for the average $\alpha_{fit}$ value. In other words, this resulted in an average $A_{fit}$ and $\alpha_{fit}$ value that would not have resulted together from a single individual fit, making the calculated diffraction pattern from the average values appear brighter than the experimental intensity. This only appeared for $\ell = 0$ because values $\alpha_{fit}$ values of the other $\ell$ modes were confined to positive values. A simple example taken from the fit data is as follows:

$$\alpha_{fit-1} = 0.108 \text{ and } A_{fit-1} = 1.22 \times 10^{-3}$$

$$\alpha_{fit-2} = -0.086 \text{ and } A_{fit-2} = 9.73 \times 10^{-4}$$

The average of the two fits above is:

$$\alpha_{fit-avg} = 0.011 \text{ and } A_{fit-avg} = 1.10 \times 10^{-3}$$

In the fit data for $\ell = 0$ at 300 K, it was found for each individual fit that:

$$A = 1.144 \times 10^{-2}|\alpha| \pm 1.74\% \qquad \text{(S1)}$$

Which leads us to predict with reasonable confidence that the $A$ value associated with a $\alpha_{fit-avg}$ would be:

$$A_{fit-avg,\ expected} \approx 1.144 \times 10^{-2}\left|\alpha_{fit-avg}\right| = 1.26 \times 10^{-4}$$

However, because of the $\alpha_{fit}$ values being averaged around zero, we see an $A_{fit-avg}$ value that is about 10x larger than it would appear in a data set. This results in a calculated image with intensity that is higher than it should be, occurring as an artifact of averaging $\alpha$ values around zero. Using equation (S1), the calculated values for $A_{fit-1}$ and $A_{fit-2}$ are within 1.26% and 1.11% of their fitted values, respectively.

Calculating the average of $\alpha$ for $\ell = 0$ from the fit data and then using the average $\alpha$ value to calculate $A$ in equation (S1) would be the best representation of the experimental data in a visual image. However, we chose not to do this to keep our presentation and methods consistent for all $\ell$ modes.

It should be noted that equation (S1) was found to apply specifically to this data set and should not be used as a standard method of calculating $A$ in other data sets. We found at different $\ell$ modes and temperatures that a similar relationship exists, but the multiplying factor is dependent on the mode and temperature. In general, we found that:

$$A \propto \alpha \qquad \text{(S2)}$$

## PHI ROTATION FIT

To fit the rotation of $\varphi$ at 340K, the first 100 frames from the right-circular polarization data set of $\ell = \frac{1}{2}$ were cropped to 234 $\mu$m x 240 $\mu$m so that each diffraction intensity pattern was centered with reasonable room around the edges. Each frame from each set was resampled to 100 x 100 pixels. The pixel size was recalculated in meters for use in the fitting algorithm. The cropped image for $\ell = 1/2$ was used in a comparison function where the x and y coordinate offsets ($x_0$ and $y_0$), beam waist ($w_0$), phase rotation offset ($\varphi_0$), amplitude offset ($A_0$), and the fractional phase step ($\alpha$) were manually fit using equation (6). The parameter values from the manual fit were used as starting values in the fitting algorithm.

The first 100 frames were divided into 20 subsets, each containing 5 consecutive frames. Each subset of 5 frames was averaged. The resulting set of 20 averaged frames was used in the fitting algorithm.

The fitting algorithm fit the parameters $x_0$, $y_0$, $w_0$, $\varphi_0$, $A_0$, and $\alpha_0$, using equation (6) for each averaged frame. This provided 20 separate fits of the data. The mean and standard deviation of the 20 subset fits was used to determine the value of each parameter. The average parameter values were used to simulate the fit results in a comparison function utilizing equation (3).

The first 437 frames of the RCP data at 340K were then fit using the average fit values for $x_0$, $y_0$, $w_0$, $A_0$, and $\alpha_0$ as fixed values with only $\varphi_0$ as a variable. Each frame was then fit individually to determine the rotation of $\varphi$ between frames. The fit was stopped at frame 437 due to the fluctuations caused by the presence of multidomain walls in the sample. The fitting algorithm is unable to account for this phenomenon, so the fit was ceased.